\documentclass[10pt]{blois07}
\usepackage{graphicx}

\usepackage{cite,./mcite}

\newcommand{\be}{\begin{equation}}
\newcommand{\ee}{\end{equation}}
\newcommand{\beeq}{\begin{eqnarray}}
\newcommand{\eeeq}{\end{eqnarray}}

\def\Pom{{\bf I\!P}}

\newcommand{\bp}{\mbox{\boldmath $p$}}

\newcommand{\bM}{\mbox{\boldmath $M$}}
\newcommand{\ket}[1]{| {#1} \rangle}
\newcommand{\bra}[1]{\langle {#1} |}

\begin{document}
\title{Diffractive production of quarkonia}
\author{Antoni Szczurek$^{1,2}$\protect\footnote{talk presented at EDS07}~,
}
\institute{$^1$Institute of Nuclear Physics Polish Academy of 
Sciences, Cracow, Poland\\ 
$^2$Institute of Physics, University of  Rzesz\'ow, Rzesz\'ow, Poland}
\maketitle
\begin{abstract}
I discuss two selected examples of diffractive production of quarkonia:
$p p \to p \eta' p$ and $p p \to p J/\psi p$. In the first case
I consider diffractive pQCD approach and $\gamma \gamma$ fusion,
in the second case the amplitude is linked to the amplitude of the
process for $J/\psi$ photoproduction at HERA. Absorption effects are
discussed briefly for the second reaction. 
\end{abstract}

\section{Introduction}
\label{sec:1}

Exclusive production of mesons was studied in details
only at fixed target collisons at CERN. 
At present, there is ongoing investigations at Tevatron
aiming to measure the exclusive production of both vector and scalar
quarkonia, but no result is yet publicly available. 
Only an upper limit for $\chi_c$ was given up to now \cite{CDF_limit}.

There is a long standing debate about the nature of the pomeron.
The approximate
$\sin^2(\Phi)$ ($\Phi$ is the azimuthal angle between outgoing
protons) dependence observed experimentally for $pp \to pp \eta'$
\cite{WA102} was interpreted in Ref.\cite{CS99} as due to
(vector pomeron)-(vector pomeron)-(pseudoscalar meson) coupling.
The QCD-inspired calculation for diffractive production of
pseudoscalar mesons was presented only recently in Ref.\cite{SPT07}.
Here I shall present some results from that analysis obtained
within the pQCD approach of Khoze-Martin-Ryskin (KMR) \cite{KMR}.

Recently the $J/\psi$ exclusive production in proton-proton and 
proton-antiproton collisions was suggested as a candidate
in searches for odderon exchange \cite{BMSC07}.
In order to identify the odderon exchange one has to consider all
other possible processes leading to the same final channel.
One of such processes, probably dominant, is pomeron-photon or
photon-pomeron fusion \cite{SS07}.

The diffractive photoproduction of $J/\psi$--mesons has been recently a
subject of thorough studies at HERA \cite{ZEUS_JPsi,H1_JPsi}, 
and serves to elucidate
the physics of the QCD pomeron and/or the small--$x$ gluon 
density in the proton (for a recent review and references, see
\cite{INS06}). Being charged particles, protons/antiprotons 
available at RHIC, Tevatron and LHC are a source of high energy
Weizs\"acker--Williams photons. Those photons interact with the other
nucleon. In some cases such an interaction leads to elastical (ground
state proton) production of $J/\psi$. In the approach presented here
the amplitude for the $p p \to p p J/\psi$ reaction is related to
the amplitude of the photoproduction $ \gamma p \to J/\psi p$ \cite{SS07}.
Such a method of calculating cross section is expected to be much
more precise than any QCD approach which does not refer to
the $e p$ HERA data.

\section{Diffractive production of $\eta'$}
\label{sec:2}


\begin{figure}[!h] 
 \centerline{\includegraphics[width=0.4\textwidth]{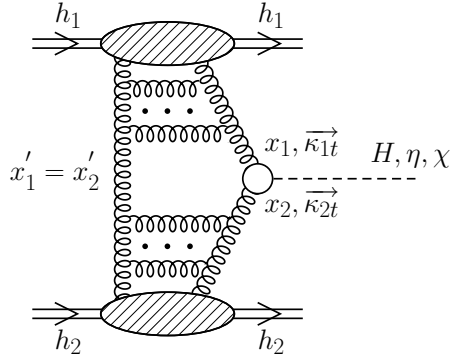}}
   \caption{\label{fig:diffraction_ugdf}
The sketch of the bare QCD mechanism. The kinematical
variables are shown in addition.}
\end{figure}


Following the formalism for the diffractive double-elastic
production of the Higgs boson one can write the amplitude
from Fig.\ref{fig:diffraction_ugdf} as
\begin{eqnarray}
{\cal M}_{pp \to p \eta' p}^{g^*g^*\to\eta'} =  i \, \pi^2 \int
d^2 k_{0,t} V(k_1, k_2, P_M) \frac{
f^{off}_{g,1}(x_1,x_1',k_{0,t}^2,k_{1,t}^2,t_1)
       f^{off}_{g,2}(x_2,x_2',k_{0,t}^2,k_{2,t}^2,t_2) }
{ k_{0,t}^2\, k_{1,t}^2\, k_{2,t}^2 } \, , 
\label{main_formula}
\end{eqnarray}
where $f's$ are skewed unintegrated gluon distributions. 
For more details see \cite{SPT07}.

\begin{figure}[!h]
\begin{center}
\includegraphics[width=0.5\textwidth]{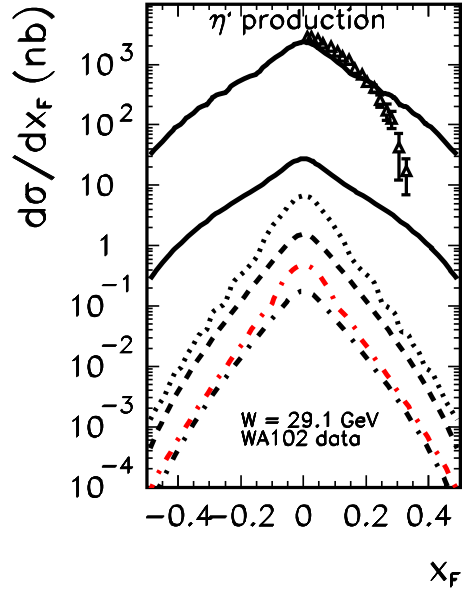}
\end{center}
\caption{$d \sigma / dx_F$ as a function of Feynman $x_F$ for W
= 29.1 GeV and for different UGDFs.
The $\gamma^* \gamma^*$ fusion contribution is shown by
the dash-dotted (red) line (second from the bottom).
The experimental data of the WA102 collaboration are shown
for comparison. The dashed line corresponds to the KL distribution, dotted
line to the GBW distribution and the dash-dotted to the BFKL distribution.
The two solid lines correspond to the Gaussian distribution with
details explained in the original paper.
No absorption corrections were included here.
}
\label{fig:dsig_dxF}
\end{figure}

As an example in Fig.~\ref{fig:dsig_dxF} I show the results of
calculations obtained with several models of UGDF
(for details see \cite{SPT07}) for relatively low energy W = 29.1 GeV.
For comparison I show also the contribution of the
$\gamma^* \gamma^*$ fusion mechanism.
The contribution of the last mechanism is much smaller than
the contribution of the diffractive QCD mechanism.


\begin{figure}[!h]     
\begin{center}
\includegraphics[width=0.35\textwidth]{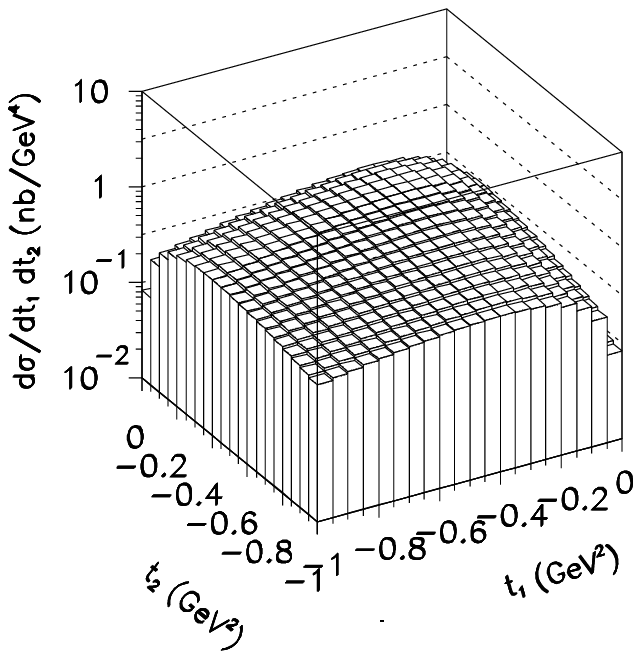}
\includegraphics[width=0.35\textwidth]{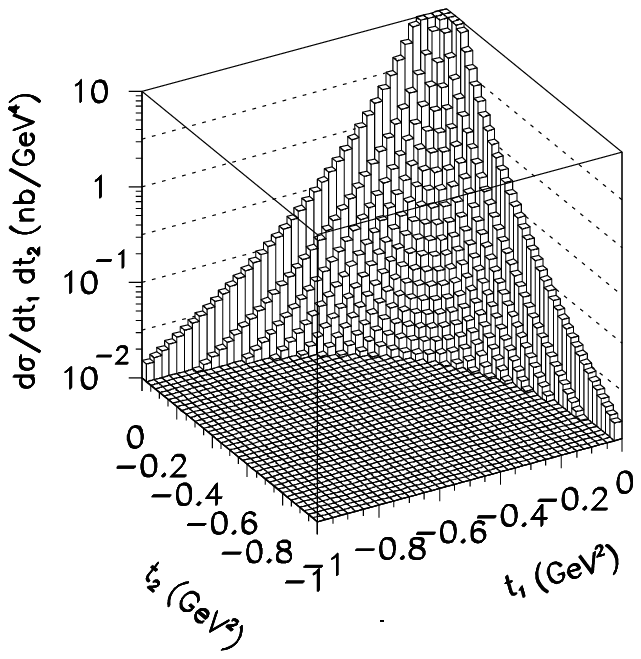}
\end{center}
\caption{Two-dimensional distribution in $t_1 \times t_2$
for the diffractive QCD mechanism (left panel), calculated with the KL
UGDF, and the $\gamma^* \gamma^*$ fusion (right panel) at
the Tevatron energy W = 1960 GeV.
No absorption corrections were included here.}
\label{fig:map_t1t2}
\end{figure}

The diffractive and $\gamma^* \gamma^*$ contributions have very
different dependence on four-momentum transfers.
In Fig.\ref{fig:map_t1t2} I present two-dimensional maps
$t_1 \times t_2$ of the cross section for the QCD mechanism (KL UGDF)
and the QED mechanism (Dirac terms only) for the Tevatron energy W = 1960 GeV.
If $ |t_1|, |t_2| > $ 0.5 GeV$^2$ the QED mechanism is clearly negligible.
However, at $|t_1|, |t_2| < $ 0.2 GeV$^2$ the QED mechanism may become
equally important or even dominant. However, the details depend
strongly on UGDFs.


\begin{figure}[!h]    
 \centerline{\includegraphics[width=0.60\textwidth]{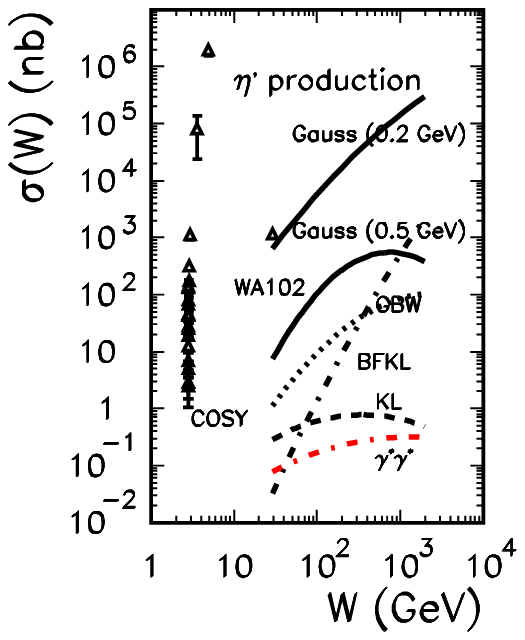}}
   \caption{ \label{fig:sig_tot_w}
$\sigma_{tot}$ as a function of center of mass energy
for different UGDFs.
The $\gamma^* \gamma^*$ fusion contribution is shown by the dash-dotted
(red) line. The world experimental data are shown for reference.
No absorption corrections were included here.}
\end{figure}

Finally in Fig.\ref{fig:sig_tot_w} I show energy dependence of
the total cross section for
the $p p \to p \eta' p$ reaction for different UGDFs.
Quite different results are obtained for different UGDFs.
The cross section with the Kharzeev-Levin type distribution (based
on the idea of gluon saturation) gives
the cross section which is relatively small and almost idependent of
beam energy.
In contrast, the BFKL distribution leads to strong energy dependence.
The sensitivity to the transverse momenta of initial gluons
can be seen by comparison of the two solid lines calculated with
the Gaussian UGDF with different smearing parameter
$\sigma_0$ = 0.2 and 0.5 GeV.
The contribution of the $\gamma^* \gamma^*$ fusion mechanism
(red dash-dotted line) is fairly small and only slowly energy dependent.


\section{Photoproduction of $J/\psi$}
\label{sec:3}


\begin{figure}[!h]    %
\begin{center}
\includegraphics[width=0.3\textwidth]{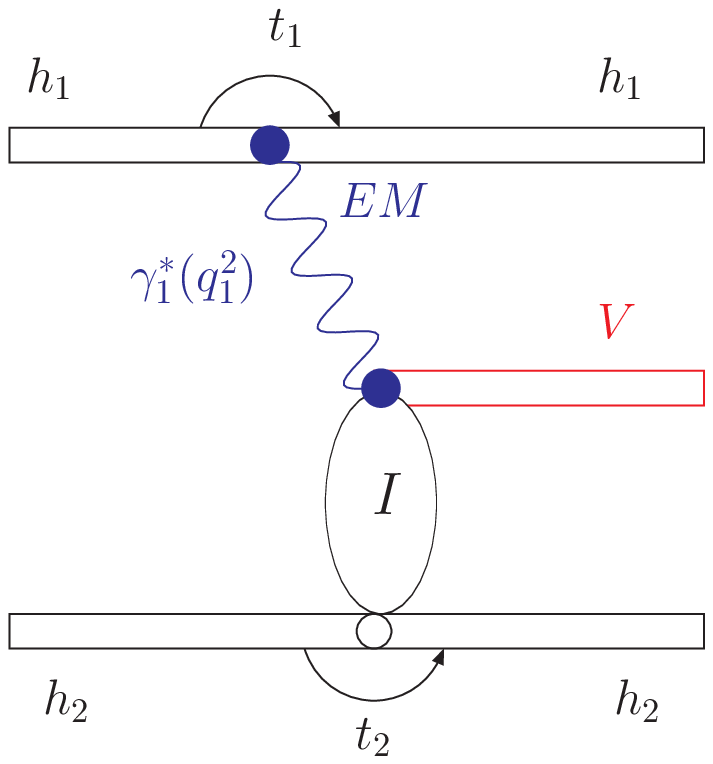}
\includegraphics[width=0.3\textwidth]{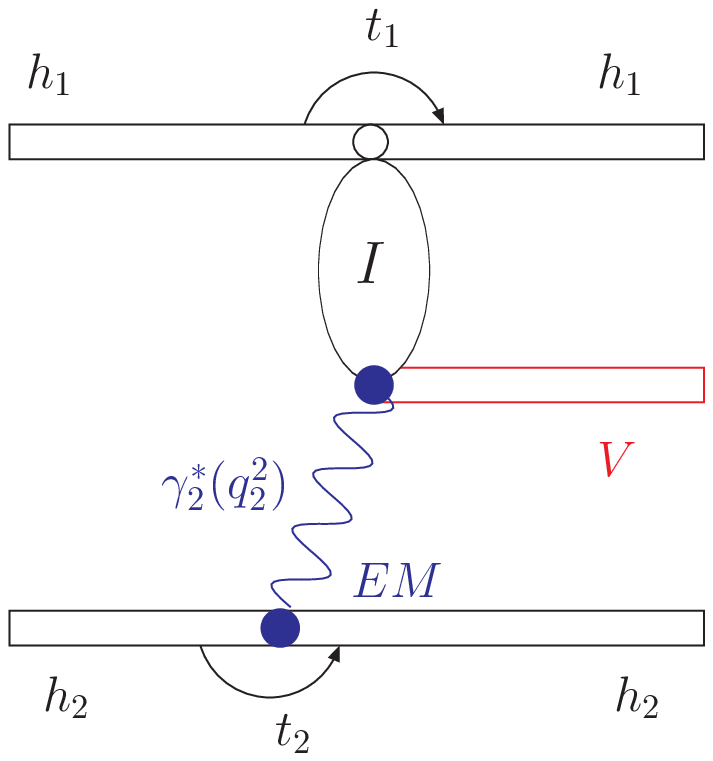}
\end{center}
   \caption{\label{fig:diagram_photon_pomeron}
The sketch of the two mechanisms considered in the present paper:
photon-pomeron (left) and pomeron-photon (right).
Some kinematical variables are shown in addition.}
\end{figure}


The basic mechanisms leading to the exclusive production
of $J/\psi$ are shown in Fig.\ref{fig:diagram_photon_pomeron}. 
The amplitude for the corresponding $2 \to 3$ process 
can be written as 
\begin{eqnarray}
{\cal M}_{h_1 h_2 \to h_1 h_2 V}^
{\lambda_1 \lambda_2 \to \lambda'_1 \lambda'_2 \lambda_V}(s,s_1,s_2,t_1,t_2) &&= 
{\cal M}_{\gamma \Pom} + {\cal M}_{\Pom \gamma} \nonumber \\
&&= \bra{p_1', \lambda_1'} J_\mu \ket{p_1, \lambda_1} 
\epsilon_{\mu}^*(q_1,\lambda_V) {\sqrt{ 4 \pi \alpha_{em}} \over t_1}
{\cal M}_{\gamma^* h_2 \to V h_2}^{\lambda_{\gamma^*} \lambda_2 \to \lambda_V \lambda_2}
(s_2,t_2,Q_1^2)   \nonumber \\
&& + \bra{p_2', \lambda_2'} J_\mu \ket{p_2, \lambda_2} 
\epsilon_{\mu}^*(q_2,\lambda_V)  {\sqrt{ 4 \pi \alpha_{em}} \over t_2}
{\cal M}_{\gamma^* h_1 \to V h_1}^{\lambda_{\gamma^*} \lambda_1 \to \lambda_V \lambda_1}
(s_1,t_1,Q_2^2)  \, . \nonumber \\
\label{Two_to_Three}
\end{eqnarray}
After some algebra it can be written in the compact form:
\begin{eqnarray} 
\bM(\bp_1,\bp_2) &&= e_1 {2 \over z_1} {\bp_1 \over t_1} 
{\cal{F}}_{\lambda_1' \lambda_1}(\bp_1,t_1)
{\cal {M}}_{\gamma^* h_2 \to V h_2}(s_2,t_2,Q_1^2)   
\nonumber \\
&&
+ e_2 {2 \over z_2} {\bp_2 \over t_2} {\cal{F}}_{\lambda_2' \lambda_2}(\bp_2,t_2)
{\cal {M}}_{\gamma^* h_1 \to V h_1}(s_1,t_1,Q_2^2)  \, . 
\end{eqnarray}
The differential cross section is given in terms of $\bM$ as
\begin{equation}
d \sigma = { 1 \over 512 \pi^4 s^2 } | \bM |^2 \, dy dt_1 dt_2
d\phi \, ,
\end{equation}
where $y$ is the rapidity of the 
vector meson, and $\phi$ is the angle between $\bp_1$ and $\bp_2$.
Notice that the interference between the two mechanisms $\gamma \Pom$
and $\Pom \gamma$ is proportional to $e_1 e_2 (\bp_1 \cdot \bp_2)$ 
and introduces a charge asymmetry as well as an angular correlation
between the outgoing protons.

\begin{figure}[!h]   
 \centerline{\includegraphics[width=0.5\textwidth]{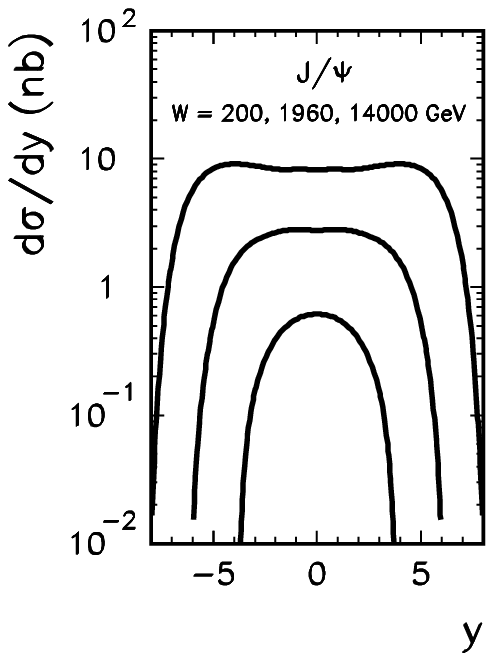}}
   \caption{ \label{fig:dsig_dy_energy}
$d \sigma / dy$ for exclusive $J/\psi$ production
as a function of $y$ for RHIC, Tevatron and LHC energies.
No absorption corrections were included here.}
\end{figure}

In Fig.\ref{fig:dsig_dy_energy} I collect rapidity distributions
for different energies relevant at RHIC, Tevatron and LHC. One observes an
occurence of a small dip in the distribution at midrapidities at LHC energy.
One should remember, however, that the distribution for the LHC energy is
long-distance extrapolation of the $\gamma^*p \to J/\psi p$
amplitude (or cross section) to unexplored yet experimentally energies
$W_{\gamma p}$.
Therefore a real experiment at Tevatron and LHC would help to constrain 
cross sections for $\gamma p \to J/\psi p$ process.


\begin{figure}[!h]   
\includegraphics[width=0.4\textwidth]{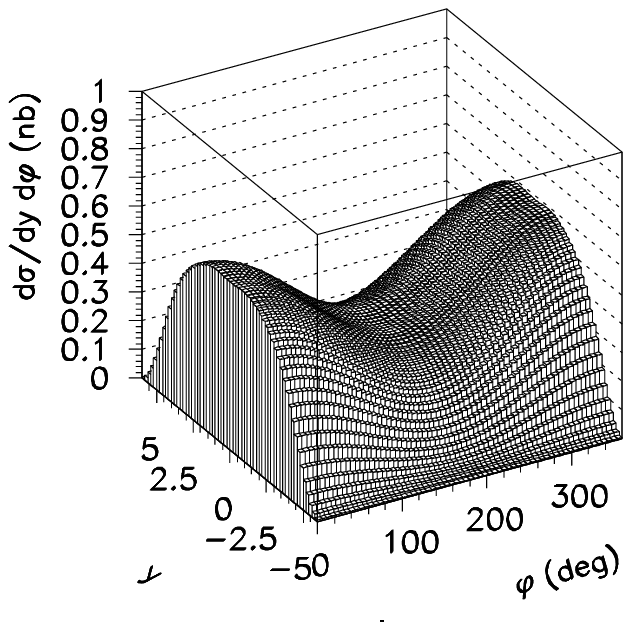}
\includegraphics[width=0.4\textwidth]{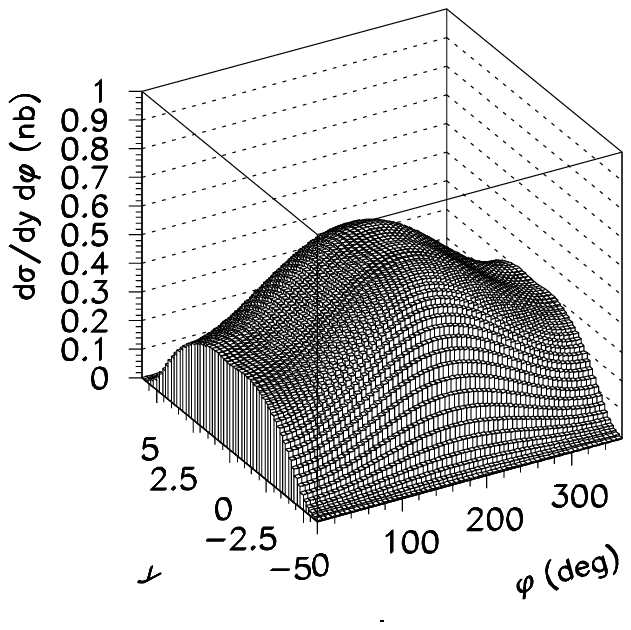} 
   \caption{ \label{fig:map_yphi.eps}
$d \sigma / dy d \Phi$ for W = 1960 GeV and for $p \bar p$ (left panel) 
and $p p$ (right panel) collisions.
No absorption corrections were included here.}
\end{figure}

In Fig.\ref{fig:map_yphi.eps} I show two-dimensional distributions
in rapidity and the azimuthal angle. 
Surprisingly, the interference effect between both diagrams
is significant over broad range of $J/\psi$ rapidity.
One can see that even at large
$J/\psi$ rapidities one observes ansisotropic distributions in
the azimuthal angle. This means that interference between photon-pomeron
and pomeron-photon mechanisms survives up to large rapidities.


\begin{figure}[!h]    %
\begin{center}
\includegraphics[width=0.3\textwidth]{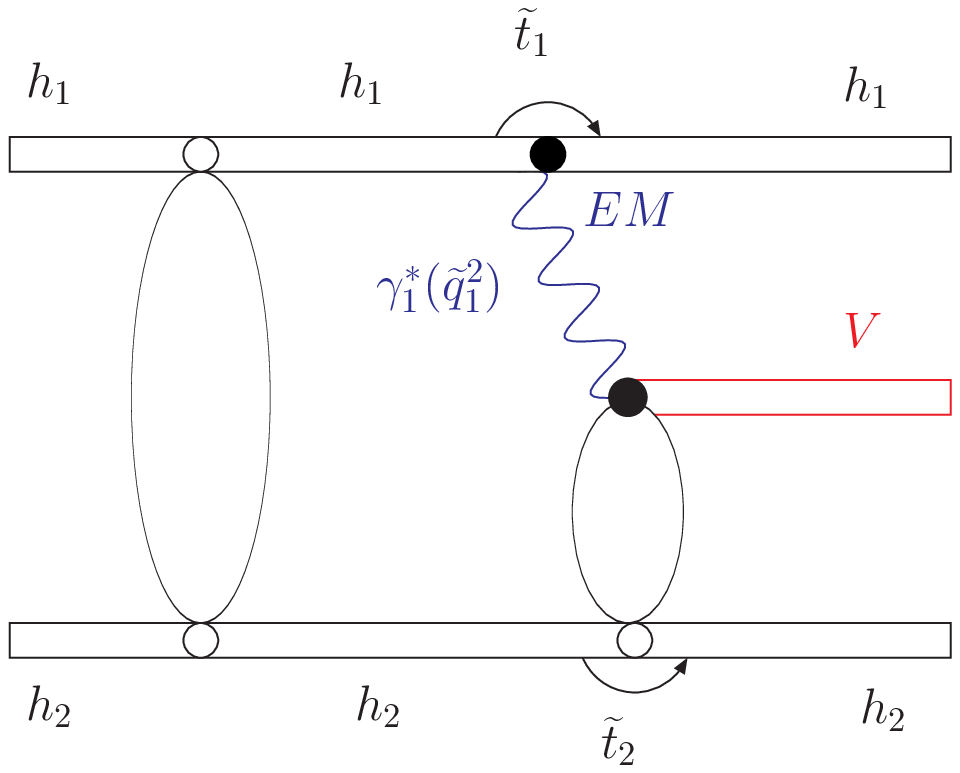}
\includegraphics[width=0.3\textwidth]{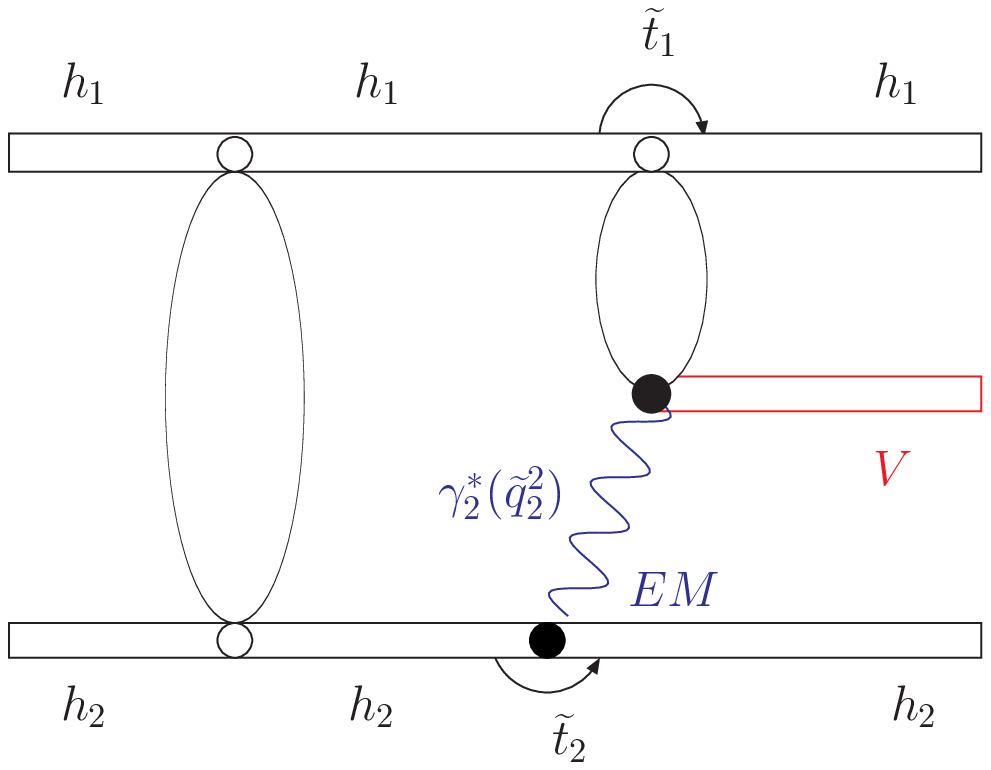}
\end{center}
   \caption{\label{fig:diagram_rescattering}
The sketch of the elastic rescattering amplitudes.
Some kinematical variables are shown in addition.}
\end{figure}

The parametrization of the $\gamma^* p \to V p$ amplitude which
describes corresponding experimental data (see \cite{SS07}) includes
effectively absorption effects due to final state $Vp$ interactions.
In the $p p \to p p J/\psi$ ($p \bar p \to p \bar p J/\psi$)
reaction the situation is more complicated as here $pp$ (or $p \bar p$)
strong rescatterings occur in addition. In Ref.\cite{SS07} we have
included only elastic rescatterings shown schematically in 
Fig.\ref{fig:diagram_rescattering}.


\begin{figure}[!h]   
\includegraphics[width=0.4\textwidth]{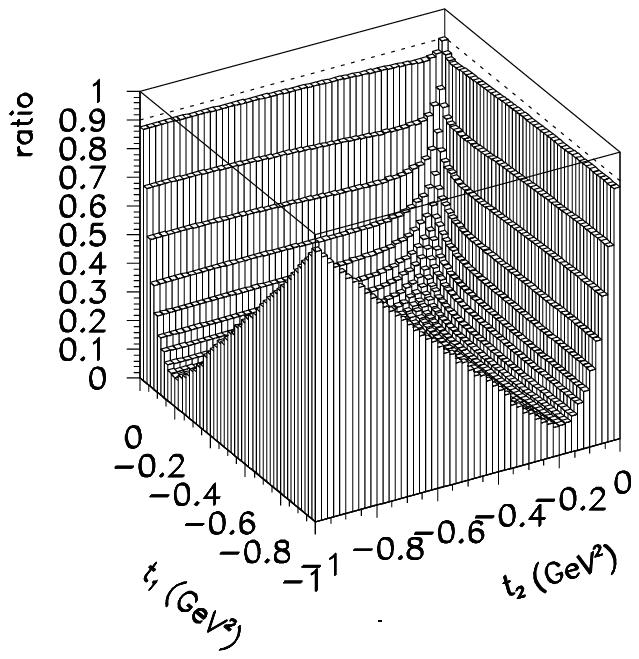}
\includegraphics[width=0.4\textwidth]{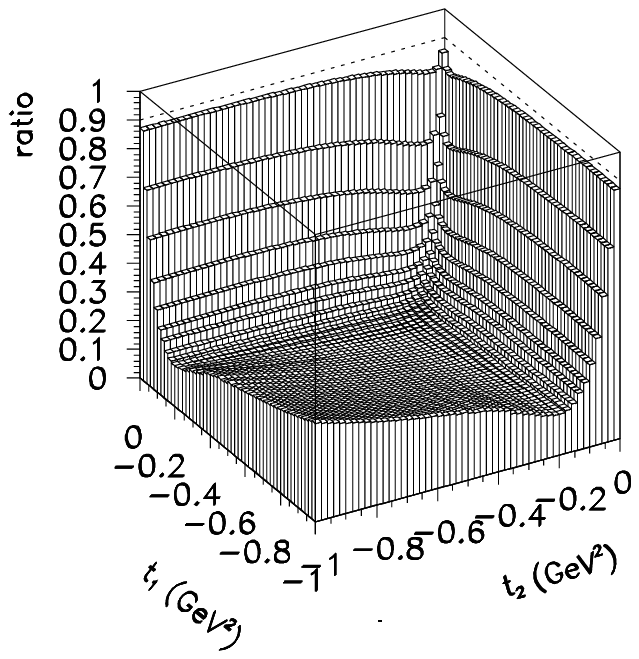}
   \caption{ \label{fig:ratio_absorption_t1t2}
The ratio of the cross sections with absorption to that without
absorption for $p \bar p$ (left panel) and $p p$ (right panel) scattering.
Here the integration over -1 GeV$^2$ $ < t_1, t_2 < $ 0.0 and
-1 $ < y < $ 1 is performed.
}
\end{figure}

In order to demonstrate the effect of the absorption
in Fig.\ref{fig:ratio_absorption_t1t2} I show
the ratio of the cross section with absorption to that without
absorption as a function of $t_1$ and $t_2$, for $p \bar p$ (left)
and $p p$ (right).
Generally, the bigger $t_1$ and/or $t_2$ the bigger the absorption.
On average, the absorption for the $p \bar p$ reaction is
smaller than the absorption for the $p p$ reactions.

\section{Summary}
\label{sec:5}

In contrast to diffractive Higgs production, in the case of ligh meson
production the main contribution to the diffractive amplitude comes
from the region of very small gluon transverse momenta and very small
longitudinal momentum fractions. In this case application of
Khoze-Martin-Ryskin UGDFs seems not justified and we have to rely
on UGDFs constructed for this region.

The existing models of UGDFs predict cross section much smaller
than the one obtained by the WA102 collaboration at the
center-of-mass energy W = 29.1 GeV. This may signal presence of
subleading reggeons at the energy of the WA102 experiment or
suggest a modificaction of UGDFs in the nonperturbative region of
very small transverse momenta.

Due to a nonlocality of the loop integral our model leads to sizeable
deviations from the $\sin^2 \Phi$ dependence (predicted in the models
of one-step fusion of two vector objects).
The $\gamma^* \gamma^*$ fusion may be of some importance only at
extremely small four-momentum transfers squared.

It was shown in \cite{SS07} that at the Tevatron energy one can study
the exclusive production of $J/\psi$ at the photon-proton center-of-mass
energies 70 GeV $ < W_{\gamma p} < $ 1500 GeV, i.e. in the unmeasured region
of energies, much larger than at HERA. At LHC this would be correspondingly
200 GeV $ < W_{\gamma p} < $ 8000 GeV. At very forward/backward
rapidities this is an order of magnitude more than possible
with presently available machines.

An interesting azimuthal-angle correlation pattern has been obtained
due to the interference of photon-pomeron and pomeron-photon 
helicity-preserving terms.

We have estimated also absorption effects. In some selected configurations
the absorption effects may lead to the occurence of diffractive minima.
The exact occurence of diffractive minima depends on the values
of the model parameters. Such minima are washed out when integrated
over the phase space or even its part. We have found that on average
the rescattering effects in proton-antiproton reactions are much bigger
than in proton-proton reactions. In this case the obvious isospin violation
is of electromagnetic origin due to the interference of diagrams
with photon exchange.

\bigskip
\centerline{ACKNOWLEDGEMENTS}
I thank Hannes Jung and his colleagues for very efficient organization
of the conference and hospitality.
The collaboration with Roman Pasechnik, Oleg Teryaev and Wolfgang Sch\"afer
on the issues presented here is acknowledged.
This work was partially supported by  the MEiN research grant~
1~P03B~028~28 (2005-08).



\end{document}